\documentclass[12pt]{article}
\usepackage{epsfig}
\title{Recent progress in QCD at low energies.}
\author{B.L.Ioffe\\
Institute of Theoretical and
Experimental Physics,\\
B.Cheremushkinskaya 25, 117218 Moscow,Russia}
\date{}
\begin{document}
\maketitle

\newcommand{\be}{\begin{equation}}
\newcommand{\ee}{\end{equation}}

\def\la{\mathrel{\mathpalette\fun <}}
\def\ga{\mathrel{\mathpalette\fun >}}
\def\fun#1#2{\lower3.6pt\vbox{\baselineskip0pt\lineskip.9pt
\ialign{$\mathsurround=0pt#1\hfil##\hfil$\crcr#2\crcr\sim\crcr}}}

\begin{abstract}

The  recent results in QCD at low energies are reported. The
theoretical analysis of hadronic $\tau$-decay is performed in
complex $q^2$-plane. The terms of perturbation theory (PT) up to
$\alpha^3_s$ are accounted, the terms of operator product
expansion (OPE) -- up do dimension 8. The momenta of charmed quark
vector current polarization operator are analyzed in 3 loops with
account of dimension 4 and 6 OPE terms. The main conclusion is
that in the range of low-energy phenomena under consideration PT
and OPE are in good agreement with experiment starting from
$Q^2=-q^2\sim 1$ GeV$^2$. The values of $\alpha_s(m^2_{\tau}),
\alpha_s\langle \overline{q}q\rangle^2$, $\langle
(\alpha_s/\pi)G^2\rangle$ and $\overline{m}_c(\overline{m}_c)$
were found.
\end{abstract}

\section{Introduction}

The first analysis of QCD at low energies in framework of PT and
OPE was performed long ago. The most powerful approach is the
method of QCD sum rules [1] (for review of old works see [2], more
recent results are reviewed in [3]). In the earlier analysis it
was accepted [1], that QCD coupling constant $\alpha_s$ is small
even at low energies $\alpha_s(1$ GeV$^2)\approx 0.3$
($\Lambda^{(3)}_{QCD}\simeq$ 100 MeV) and only first order
$\alpha_s$ corrections were accounted. Now it is clear, that at
low $Q^2\sim 1-5$ GeV$^2$ $\alpha_s(Q^2)$ is about two times
larger, then it was taken in [1]. From the other side, new, more
precise  experimental data are now available. For these reasons it
is necessary the reanalysis  of QCD at low energies with account
of higher order terms of PT and OPE in comparison with
experimental data. In this report I present the results of such
analysis.

\section{Test of QCD at low energies on the basis of
$\tau$-decay data. Determination of $\alpha_s(m^2_{\tau})$ and of
condensate values.}

Recently, collaborations ALEPH [4], OPAL [5] and CLEO [6] had
measured with a good accuracy the relative probability of hadronic
decays of $\tau$-lepton $R_{\tau} = B(\tau \to \nu_{\tau} +
hadrons)/B(\tau \to \nu_{\tau} e\overline{\nu}_e)$, the vector $V$
and axial $A$ spectral functions and the cross section of
$e^+e^-\to hadrons$. Below I present the results of the
theoretical analysis of $\tau$-decay data basing on the operator
product expansion (OPE) in QCD [7,8]. In the perturbation theory
series the terms up to $\alpha^3_s$ will be taken into account, in
OPE -- the operators up to dimension 8.

Consider the polarization operator of hadronic currents
\be
\Pi^J_{\mu \nu} = i~ \int~ e^{iqx} \langle T J_{\mu} (x) J_{\nu}
(0)^{\dag} \rangle dx = (q_{\mu} q_{\nu} - q_{\mu \nu} q^2)
\Pi^{(1)}_J (q^2) + q_{\mu} q_{\nu} \Pi^{(0)}_J (q^2), \ee $$
\mbox{where} ~~~~ J = V,A; ~~~ V_{\mu} = \bar{u} \gamma_{\mu} d,
~~~ A_{\mu} = \bar{u} \gamma_{\mu} \gamma_5 d. $$ The spectral
functions measured in $\tau$-decay are imaginary parts of
$\Pi^{(1)}_J(s)$ and $\Pi^{(0)}_J(s)$, ~ $s = q^2$

\be
v_1/a_1(s) = 2\pi Im \Pi^{(1)}_{V/A} (s + i 0), ~~~ a_0(s) = 2 \pi
Im \Pi^{(0)}_A (s + i0) \ee Functions $\Pi^{(1)}_V(q^2)$ and
$\Pi^{(0)}_A(q^2)$ are analytical functions in the $q^2$ complex
plane  with a cut along the right-hand semiaxis starting from $4
m^2_{\pi}$ for $\Pi^{(1)}_V(q^2)$ and $9m^2_{\pi}$ for
$\Pi^{(0)}_A( q^2)$. Function $\Pi^{(1)}_A(q^2)$ has kinematical
pole at $q^2 = 0$. This is a specific feature of QCD following
from chiral symmetry within massless $u$ and $d$ quarks and from
its spontaneous violation. The kinematical pole appears due to
one-pion state contribution into $\Pi_A(q)$, which has the form
[7]
\be
\Pi^A_{\mu \nu}(q)_{\pi} = -\frac{f^2_{\pi}}{q^2} (q_{\mu} q_{\nu}
- q_{\mu \nu} q^2) - \frac{m^2_{\pi}}{q^2} q_{\mu} q_{\nu}
\frac{f^2_{\pi}}{q^2 - m^2_{\pi}} \ee Consider first the ratio of
the total probability of hadronic decays of $\tau$-lepons into
states with zero strangeness to the probability of $\tau \to
\nu_{\tau} e \overline{\nu}_e$. This ratio is given by the
equality [9]
$$
R_{\tau, V+A} = \frac{B(\tau \to \nu_{\tau} + hadrons_{S=0})}{B(\tau \to
\nu_{\tau} e\bar{\nu}_e)} =
$$
\be
= 6 \vert V_{ud} \vert^2 S_{EW}~ \int\limits^{m^2_{\tau}}_{0}~
\frac{ds}{m^2_{\tau}} \Biggl ( 1 - \frac{s}{m^2_{\tau}} \Biggr )^2
\Biggl [ \Biggl ( 1 + 2 \frac{s}{m^2_{\tau}} \Biggr ) (v_1 + a_1
+a_0)(s) - 2 \frac{s}{m^2_{\tau}} a_0(s) \Biggr ] \ee where $\vert
V_{ud} \vert = 0.9735 \pm 0.0008$ is the matrix element of the
Kabayashi-Maskawa matrix, $S_{EW} = 1.0194 \pm 0.0040$ is the
electroweak correction [10]. Only one-pion state is practically
contributing to the last term in (4) and it appears to be small:
\be
\Delta R^{(0)}_{\tau} = - 24 \pi^2 \frac{f^2_{\pi} m^2_{\pi}}{m^4_{\tau}} =
- 0.008
\ee
Denote
\be
\omega(s) \equiv v_1 + a_1 +a_0 = 2\pi Im [\Pi^{(1)}_V(s) +
\Pi{(1)}_A(s) + \Pi^{(0)}_A(s) ] \equiv 2 \pi Im \Pi(s) \ee As
follows from eq.(3), $\Pi(s)$ has no kinematical pole, but only
right-hand cut. It is convenient to transform the integral in
eq.(4) into that over the circle of radius $m^2_{\tau}$ in the
complex $s$ plane [11]-[13]:
\be
R_{\tau,\, V+A} = 6\pi i |V_{ud}|^2 S_{EW}
\oint_{|s|=m_\tau^2}\!{ds\over m_\tau^2} \left( 1-{s\over
m_\tau^2} \right)^2 \left( 1+2 {s\over m_\tau^2}\right) \Pi (s) +
\Delta R_\tau^{(0)} \label{35} \ee Calculate first the
perturbative contribution into eq.(7). To this end, use the Adler
function $D(Q^2)$:
\be
D(Q^2) \, \equiv \, - 2\pi^2 \,{ d\Pi(Q^2)\over d\ln{Q^2}}
\,=\,\sum_{n\ge 0} K_n a^n
 \; , \qquad a\equiv {\alpha_s \over \pi}\; , \qquad  Q^2\equiv
 -s,\label{36} \ee
the perturbative expansion of which is known up to terms $\sim
\alpha^3_s$. In $\overline{MS}$ regularization scheme $K_0 = K_1 =
1$,~~ $K_2 = 1.64$ [14], $K_3 = 6.37$ [15] for 3 flavours and for
$K_4$ there is the estimate $K_4 = 25 \pm 25$ [16]. The
renormgroup equation yields
\be
{d a \over d \ln{Q^2}} \, =\, -\beta(a) \,=\, - \sum_{n\ge 0}
\beta_n a^{n+2}  \qquad  \Rightarrow
 \qquad \ln{Q^2\over \mu^2}\, = \,-\, \int_{a(\mu^2)}^{a(Q^2)}
 {da\over \beta(a)},
\label{37} \ee in the $\overline{MS}$ scheme for three flavours
$\beta_0 = 9/4$,~$\beta_1 = 4$, ~ $\beta_2 = 10.06$, ~ $\beta_3 =
47.23$ ~[17,18]. Integrating over eq.(8) and using eq.(19) we get
\be
\Pi(Q^2)\,=\,{1\over 2\pi^2} \int_{a(\mu^2)}^{a(Q^2)} D(a)
{da\over \beta(a)} \label{38}
\ee

Put $\mu^2 = m^2_{\tau}$ and choose some (arbitrary) value
$a(m^2_{\tau})$. With the help of eq.(9)  one may determine then
$a(Q^2)$ for any $Q^2$ and by analytical continuation for any $s$
in the complex plane. Then, calculating (10) find $\Pi(s)$ in the
whole complex plane. Substitution of $\Pi(s)$ into eq.(7)
determines $R_{\tau}$ for the given $a(m^2_{\tau})$ up to power
corrections. Thereby, knowing $R_\tau$ from experiment it is
possible to find the corresponding to it $a(m^2_\tau)$. Note, that
with such an approach there is no need to expand the nominator in
eqs.(9),(10) in the inverse powers of $ln Q^2/\mu^2$.
Particularly, there is no expansion on the right-hand semiaxis in
powers of the parameter $\pi/ln (Q^2/\mu^2)$, which is not small
in the investigated region of $Q^2$. Advantages  of transformation
of the integral over the real axis (4) in the contour integral are
the following.  It can be expected that the applicability region
of the theory presented as perturbation theory (PT) + operator
expansion (OPE) in the complex $s$-plane is off the shadowed
region in Fig.1. It is evident that at positive and comparatively
small $s$ PT+OPE do not work. At negative $s = -Q^2$ in $\alpha_s$
order a nonphysical pole appears, in higher orders, according with
(9) it is replaced by a nonphysical cut, which starts from the
point $-Q^2_0$, determined by the formula


\begin{figure}[tb]
\hspace{30mm} \epsfig{file=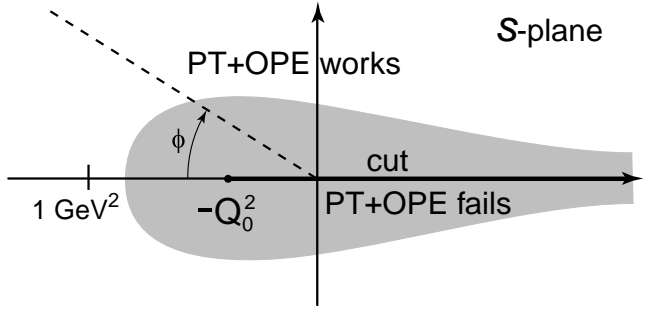, width=85mm} \caption{ The
applicability region of PT and OE in the complex plane $s$. In the
dashed region PT + OE does not work.}
\end{figure}
\be
\ln{Q_0^2\over \mu^2}\, = \,-\, \int_{a(\mu^2)}^\infty {da\over
\beta(a)}\label{39} \ee Integration over the contour allows one to
obviate the dashed region in Fig.1 (except for the vicinity of the
positive semiaxis, the contribution of which, is suppressed  by
the factor $(1 - \frac{s}{m^2_{\tau}})^2$ in eq.(7)), i.e. to work
in the applicability region of PT+OPE. The OPE terms, i.e., power
corrections to polarization operator, are given by the formula:
$$ \Pi(s)_{nonpert}=\sum_{n\ge 2} {\left<O_{2n}\right>\over
(-s)^n} \left( 1+ c_n {\alpha_s\over \pi} \right)=
 {\alpha_s\over 6 \pi\, Q^4} \left< G_{\mu\nu}^aG_{\mu\nu}^a\right>\left( 1+
{7\over 6} {\alpha_s\over \pi} \right)+$$
\be
+  {128\over 81\, Q^6}\, \pi\alpha_s \left<\bar{q}q\right>^2
\left[ 1 + \left({29\over 24} + {17\over 18}\ln{Q^2\over \mu^2}
\right){\alpha_s\over \pi}
 \right] + {\left<O_8\right>\over Q^8}  \label{40}
\ee ($\alpha_s$-corrections to the 1-st and 2-d terms in eq.(11)
were calculated in [19] and [20], respectively). Contributions of
the operator with $d=2$ proportional to $m^2_u$, $m^2_d$ and of
the condensate $2(m_u + m_d)\langle 0 \vert \bar{q}q \vert 0
\rangle$ are neglected. (The latter is of an order of magnitude
smaller than the gluonic condensate contribution). When
calculating the d=6  term, factorization hypothesis was used. It
can be readily seen that d=4 condensates (up to small $\alpha_s$
corrections) give no contribution into the integral over contour
eq.(7). The contribution from the condensate $\langle O_8 \rangle$
may be estimated as $\vert \langle O_8 \rangle \vert < 10^{-3}
GeV^8$ and appears to be negligibly small. $R_{\tau,V+A}$ may be
represented as
$$ R_{\tau, V+A}  = 3|V_{ud}|^2 S_{EW}\left(
\,1\,+\,\delta_{EW}'\,+\,\delta^{(0)}\,+\,\delta^{(6)}_{V+A} \,
\right) +\Delta R^{(0)}~=$$
\be
= ~3.475 \pm 0.022 \label{41} \ee where $\delta^{\prime}_{em} =
(5/12 \pi)\alpha_{em}(m^2_{\tau}) = 0.001$ is electromagnetic
correction [21], $\delta^{(6)}_{A+V} = -(5\pm2)\cdot 10^{-3}$ is
the contribution of d=6 condensate (see below) and $\delta^{(0)}$
is the PT correction. The right-hand part presents the
experimental value obtained as a difference between the total
probability of hadronic decays $R_{\tau} = 3.636 \pm 0.021$ [22]
and the probability of decays in states with the strangeness $S =
-1 ~~R_{\tau,s} = 0.161 \pm 0.007$ [23,24].  For perturbative
correction it follows from eq.(13)
\be \delta^{(0)} = 0.206 \pm 0.010 \label{42}\ee Employing the
above described method in ref.[8] the constant
$\alpha_s(m^2_{\tau})$  was found from (14)
\be
\alpha_s(m^2_{\tau}) = 0.355 \pm 0.025  \label{43}\ee The
calculation was made with the account of terms $\sim
\alpha^3_{\tau}$, the estimate of the effect of the terms
$\sim\alpha^4_s$ is accounted for in the error. May be, the error
is underestimated (by 0.010-0.015), since the theoretical and
experimental errors were added in quadratures.

I determine now the values of condensates basing on the data
[3]-[4] on spectral functions. It is convenient first to consider
the difference $\Pi_V-\Pi_A$, which is not contributed by
perturbative terms and there remains only the OPE contribution:
\be
\label{ope1} \Pi_V^{(1)}(s)-\Pi_A^{(1)}(s)\,=\,\sum_{D\ge 4}
\,{O^{V-A}_D \over (-s)^{D/2} } \left( \,1\,+\,c_D {\alpha_s\over
\pi}\, \right)\,\label{44} \ee The gluonic condensates
contribution falls out in the $V-A$ difference and only the
following condensates with d=4,6,8 remain
\be O^{V-A}_4  =  2 \,(m_u +m_d)\,<\bar{q}q>  \; = \; -\, f_\pi^2
m_\pi^2 \label{45}\ee
$$ O^{V-A}_6  =  2\pi \alpha_s \left<\,
(\bar{u}\gamma_\mu\lambda^a d)(\bar{d}\gamma_\mu \lambda^a u) -
(\bar{u}\gamma_5\gamma_\mu\lambda^a d)(\bar{d}\gamma_5\gamma_\mu
\lambda^a u)\, \right>  = $$ \be = -\,{64\pi\alpha_s\over 9}
<\bar{q}q>^2 \label{46} \ee
\be
 O^{V-A}_8  =   8\pi \alpha_s \, m_0^2
<\bar{q}q>^2 \;,~~-q\langle0\mid
\overline{q}\sigma_{\mu\nu}\frac{\lambda^n}{2}G^n_{\mu\nu} q\mid
0\rangle \equiv m^2_0\langle 0\mid \overline{q}q\mid 0\rangle
\label{47} \ee In the right-hand of (18) and the first of eq.'s
(19) the factorization hypothesis was used. Calculation of the
coefficients at $\alpha_s$ in eq.(16) gave $c_4 = 4/3$ [19] and
$c_6 = 89/48$ [20]. The value of $\alpha_s(m^2_{\tau})$  (15)
corresponds to $\alpha_s(1 GeV^2) = 0.60$. Thus, if we take for
quark condensate at the normlization point $\mu^2 = 1 GeV^2$ the
value following from Gell-Mann-Oakes-Renner relation at
$m_u+m_d=11.7$ MeV, then vacuum condensates with the account of
$\alpha_s$-corrections appear to be equal (at $\mu^2 = 1 GeV^2$):
\be
O_4=-4.22 \cdot 10^{-4}~GeV^4 \label{48}\ee
\be
O_6=-3.75 \cdot 10^{-3}~GeV^6 \label{49}\ee
\be
O_8=2.5 \cdot 10^{-3}~GeV^8 \label{50}\ee (In what follows,
indeces $V-A$ will be omitted and $O_D$ will mean condensates with
the account of $\alpha_s$ corrections).

Our aim is to compare OPE theoretical predictions with
experimental data on $V-A$ structure functions measured in
$\tau$-decay and the values of $O_6$ and $O_8$ found from
experiment to compare with eqs.(21),(22).  Numerical values of
$O_6$ and $O_8$ (21),(22) do not strongly differ. This indicates
that OPE asymptotic series (16) at $Q^2 = -s \sim 1 GeV^2$
converge badly and, may be, even diverge and the role of higher
dimension operators may be essential.  Therefore it is necessary
to improve the series convergence. The most plausible method is to
use Borel transformation. Write for $\Pi^{(1)}_V - \Pi^{(1)}_A$
the subtractionless dispersion relation
\be
\Pi^{(1)}_V(s)-\Pi^{(1)}_A(s)\,=\,{1\over 2\pi^2}\int_0^\infty
{v_1(t)-a_1(t)\over t-s} \, dt\,+ \,{f_\pi^2\over s} \label{51}
\ee  Put $s = s_0^{i \phi}$ ($\phi = 0$ on the upper edge of the
cut) and make the Borel transformation in $s_0$. As a result, we
get the following sum rules for the real and imaginary parts of
(23):
\be
\int_0^\infty
\exp{\!\left({s\over M^2}\cos{\phi}\right)}\cos{\!\left({s\over
M^2}\sin{\phi}\right)} (v_1-a_1)(s)\,{ds\over 2\pi^2} \, = \,
f_\pi^2+\,\sum_{k=1}^\infty (-)^k {\cos{(k\phi)} \,O_{2k+2}\over k!\,
M^{2k}} \label{52}\ee
\be
\int_0^\infty \exp{\!\left({s\over
M^2}\cos{\phi}\right)}\,\sin{\!\left({s\over
M^2}\sin{\phi}\right)} (v_1-a_1)(s)\,{ds\over 2\pi^2 M^2} \, =
\,\sum_{k=1}^\infty (-)^k {\sin{(k\phi)} \,O_{2k+2}\over k!\,
M^{2k+2}} \label{53}\ee The use of the Borel transformation along
the rays in the complex plane has a number of advantages. The
exponent index is negative at $\pi/2 < \phi < 3 \pi/2$. Choose
$\phi$ in the region $\pi/2 < \phi < \pi$. In this region, on one
hand, the shadowed area in fig.1 in the integrals (24),(25) is
touched to a less degree, and on the other hand, the contribution
of large $s$, particularly, $s > m^2_{\tau}$ , where experimental
data are absent, is exponentially suppressed. At definite values
of $\phi$ the contribution of some condensates vanishes, what may
be also used.  In particular, the condensate $O_8$ does not
contribute to (24) at $\phi = 5 \pi/6$ and to (25) at $\phi = 2
\pi/3$, while  the contribution of $O_6$ to (24) vanishes at $\phi
= 3 \pi/4$. Finally, a well known advantage of the Borel sum rules
is factorial suppression of higher dimension terms of OPE.
Figs.2,3 presents the results of the calculations of left-hand
parts of eqs.(24),(25) on the basis of the ALEPH [1] experimental
data comparing with OPE predictions -- the right-hand part of
these equations.


\begin{figure}[tb]
\hspace{0mm} \epsfig{file=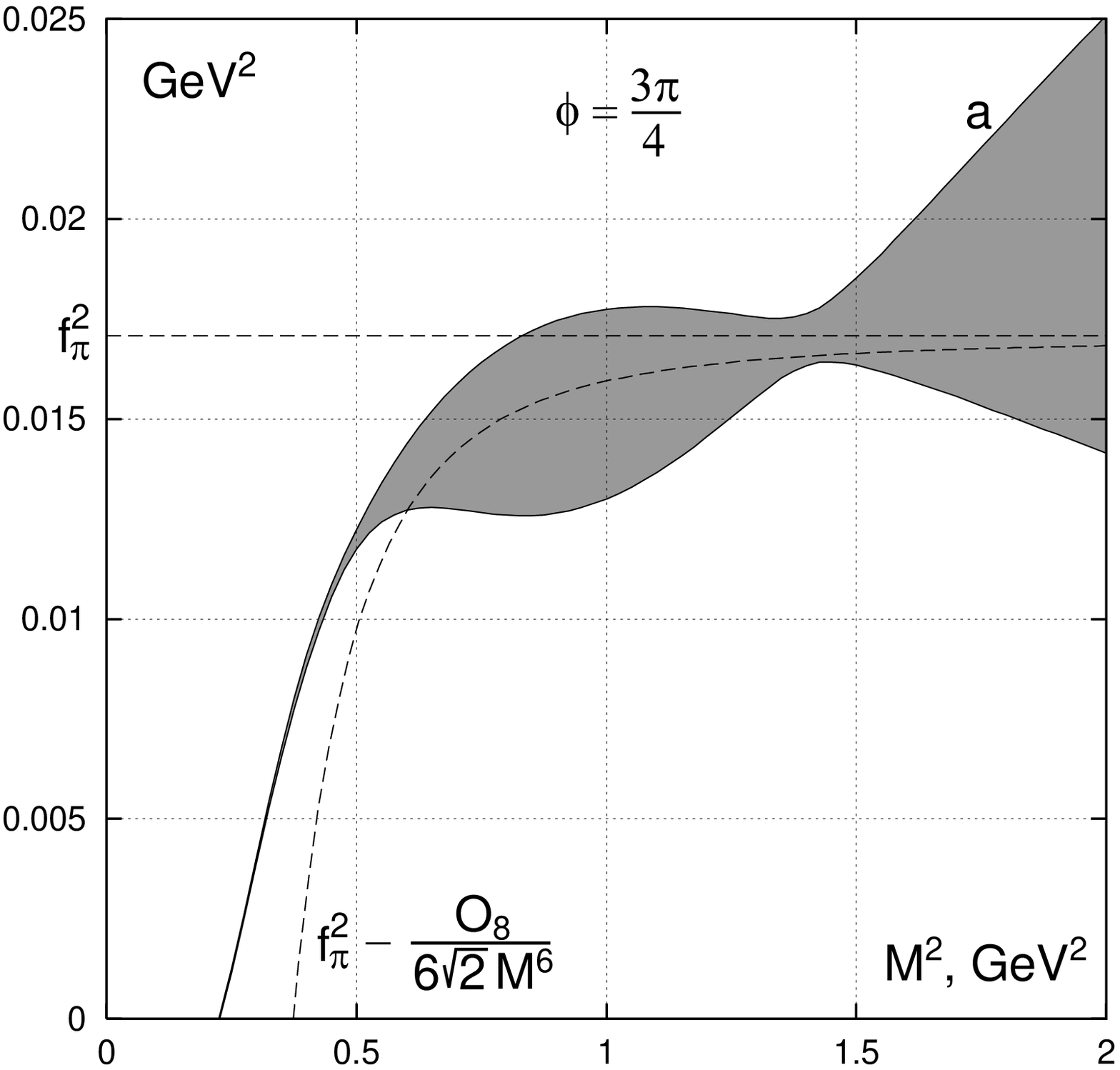, width=70mm} \hspace{5mm}
\epsfig{file=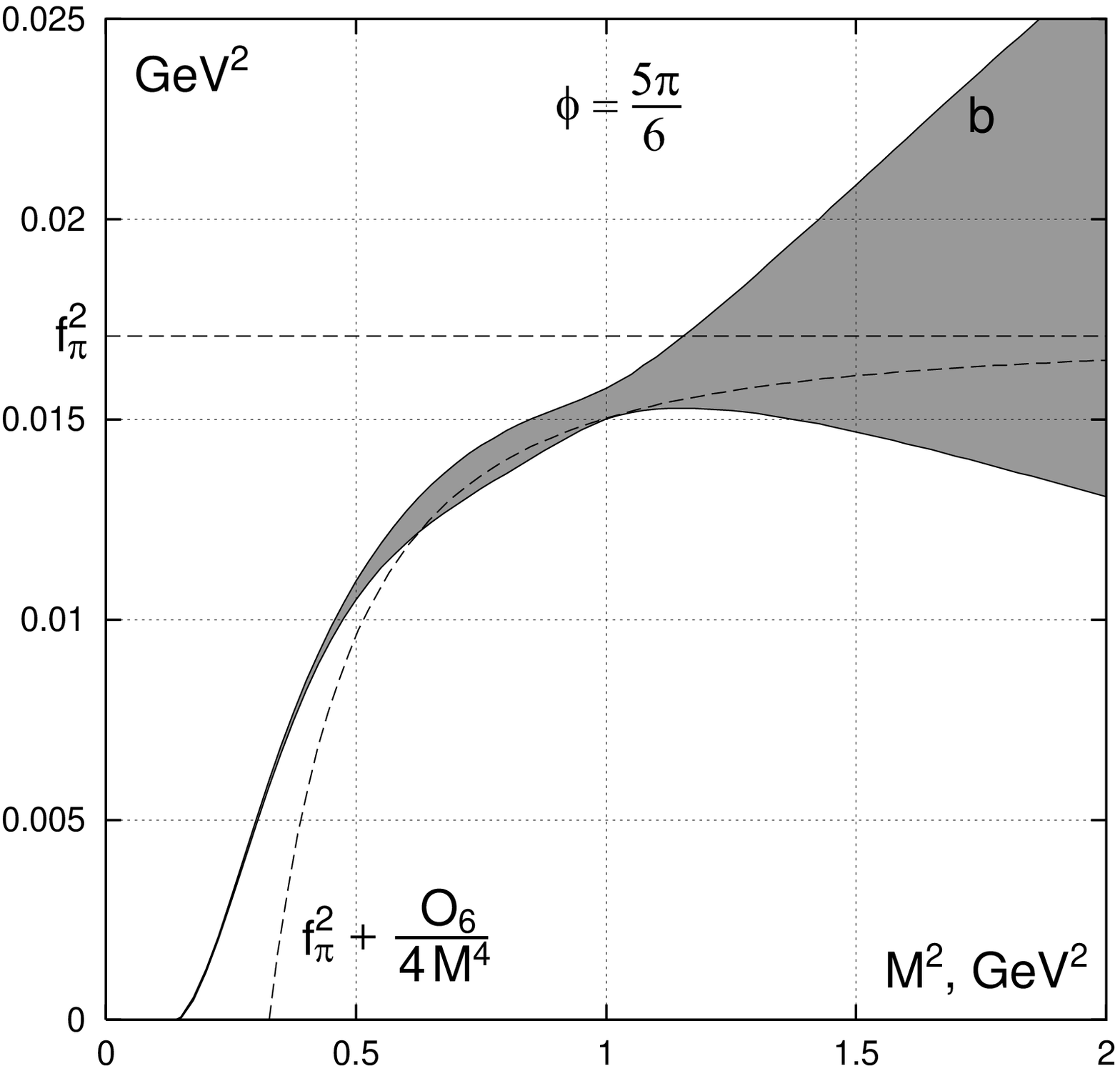, width=70mm} \caption{Eq.(24): the
left-hand part is obtained basing on the experimental data, the
shaded region corresponds to experimental errors; the right-hand
part -- the theoretical one -- is represented by the dotted curve,
numerical values of condensates are taken to be equal to the
central values of eqs.(26),(27); a) $\phi = 3 \pi/4$,~ b) $\phi =
5 \pi/6$.}
\end{figure}
\begin{figure}[tb]
\hspace{0mm} \epsfig{file=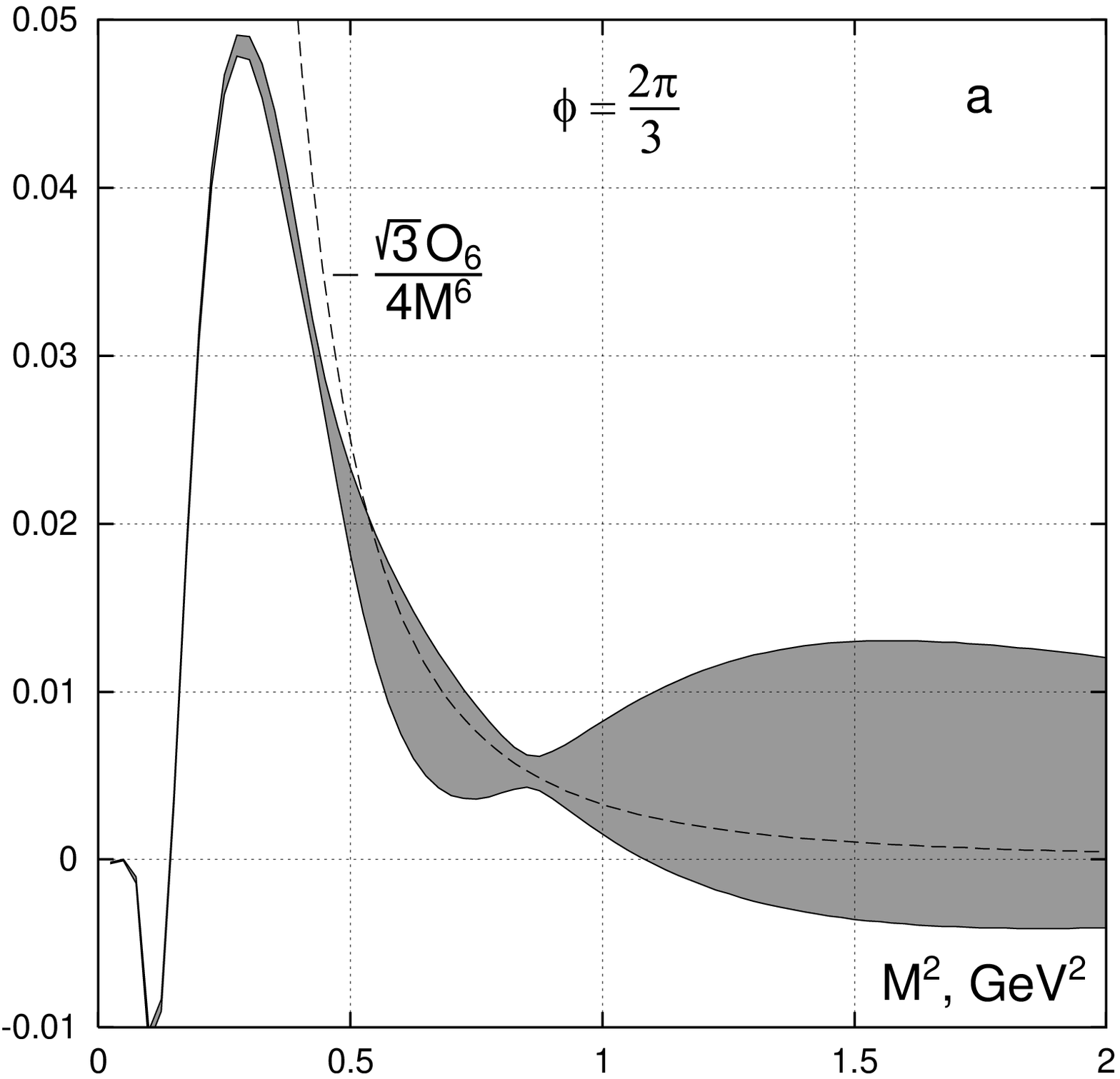, width=70mm} \hspace{5mm}
\epsfig{file=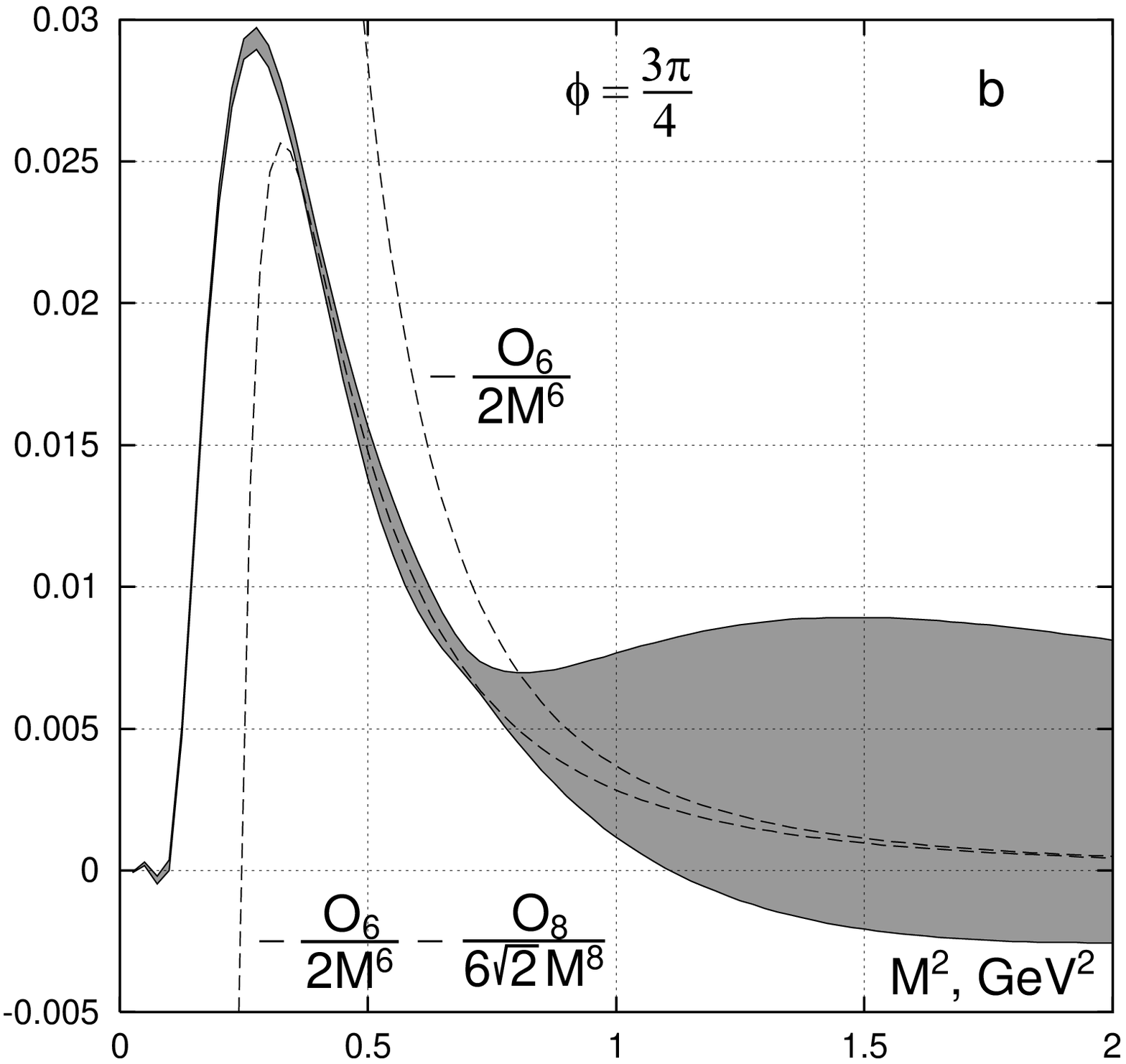, width=70mm} \caption{The same for
eq.(53): a) $\phi = 2 \pi/3$, ~ b) $\phi = 3 \pi/4$. }
\end{figure}
The experimental data are best described at the values [7]
\be
O_6=-(6.8\pm 2.1)\cdot 10^{-3}~GeV^6 \ee
\be
O_8=-(7\pm 4)\cdot 10^{-3}~GeV^8\label{55}\ee When estimating
errors in (26),(27), an uncertainty of higher dimension operator
contribution was taken into account in addition to experimental
errors. (For details -- see [7]).

As is seen from the figures, at these values of condensates a good
agreemeent with experiment starts rather early -- at $M^2
> 0.5 GeV^2$.
The values (26),(27) are by a factor of 1.5-2 larger than
(21),(22). As was discussed above, the accuracy of (21),(22) is of
order 50$\%$. Therefore, the most plausible is that the real value
of condensates $O_6,O_8$ is somewhere close to the lower edge of
errors in (26),(27).

Consider now the polarization operator $\Pi(s)$ defined in (6) and
condensates entering OPE for $\Pi(s)$ (see (12)). In principle,
the perturbative terms contribute to chirality conserving
condensates. If we will follow the separation method of
perturbative and nonperturbative contribution by introducing
infrared cut-off [25,26], then such a contribution would really
appear due to the region of virtualities smaller than $\mu^2$.  In
the present paper, according to [8],
 an another method is exploited, when the $\beta$-function is expanded only
in the number of loops, (see eq.(10) and the text after it) but
not in $1/lnQ^2$.  So, the dependence of condensates on the
normalization point $\mu^2$ is determined only by perturbative
corrections, as is seen in (12). Condensates determined in such a
way may be called $n$-loop ones (in the given case -- 3-loop).
Consider the Borel transformation of the sum $\Pi(s)_{pert} +
\Pi(s)_{nonpert}$ where $\Pi(s)_{pert}$ is given by eq.(10), and
$\Pi(s)_{nonpert}$ -- by eq.(12).  Fig.4 presents the results of
3-loop calculation for two values of $\alpha_s(m^2_{\tau})$ --
0.355 and 0.330. The widths of the bands correspond to theoretical
error taken to be equal to the last accounted term $K_3a^2$ in the
Adler function (8). (The same result for the error is obtained if
one takes 4 loops in $\beta$-function and puts $K_4 = 50 \pm 50$).
The dotted line corresponds to the sum  of contributions of
gluonic condensate $\langle \frac{\alpha_s}{\pi}G^2\rangle =
0.012$ GeV$^4$ and $O^{V+A}_6$ condensate in (12) with numerical
value corresponding to $O^{V-A}_6$ (26). The dots with errors
present experimental data. (The contribution of the operators  $d
= 4$ and $d = 6$ is given separately in the insert).

It is seen that the curve with $\alpha_s(m^2_{\tau}) = 0.330$ and
condensate contributions can be agreed with experiment, starting
from $M^2 = 1.1 GeV^2$, the agreement being improved at smaller
values $\langle 0 \vert \frac{\alpha_s}{\pi} G^2 \vert 0 \rangle$
than 0.012 GeV$^4$. The curve with $\alpha_s(m^2_{\tau}) = 0.355$
with the account of condensates coincides with experiment only at
$M^2
> 1.5 GeV^2$. The same tendency persist for the Borel sum rules
taken along the rays in the $s$ complex plane at various $\phi$.
Fig.5 gives the sum rule for $\phi = 5 \pi/6$. From consideration
of this and of other sum rules there follows the estimation for
gluonic condensate:


\begin{figure}[tb]
\hspace{30mm} \epsfig{file=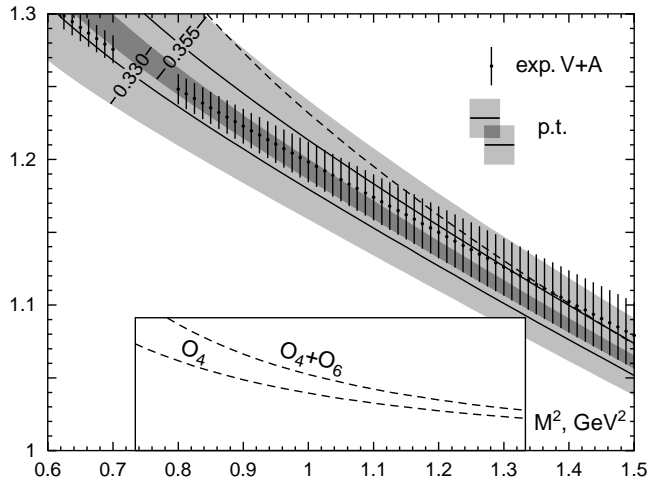, width=85mm} \caption{The
results of the Borel transformation of $V+A$ correlator for two
values $\alpha_s(m^2_{\tau} = 0.355$ and $\alpha_s(m^2_{\tau}) =
0.330$. The widths of the bands correspond to PT errors, dots with
errors -- experimental data. The dotted curve is the sum of the
perturbative contribution at $\alpha_s(m^2_{\tau}) = 0.330$ and
$O_4$,~ $O_6$ condensates.}
\end{figure}


\begin{figure}[tb]
\hspace{30mm} \epsfig{file=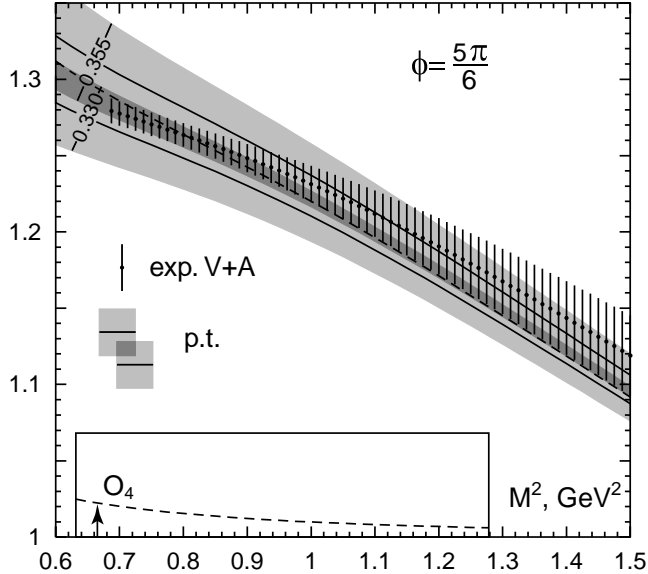, width=85mm} \caption{ The
result of the Borel transformation along the ray at $\phi = 5
\pi/6$. The dotted line corresponds to the central value of
gluonic condensate (56), added to the PT curve with
$\alpha_s(m^2_{\tau}) = 0.330$. }
\end{figure}
\be
\langle 0\mid \frac{\alpha_s}{\pi} G^a_{\mu\nu} G^a_{\mu\nu} \mid
0 \rangle = 0.006 \pm 0.012 ~GeV^4\label{56}\ee The best agreement
of the theory with experiment in the low $Q^2$ region (up to $\sim
2\%$  at $M^2 > 0.8 GeV^2$) is obtained at $\alpha_s(m^2_{\tau}) =
0.330$ which corresponds to $\alpha_s(m^2_z) = 0.118$.

 It was shown [8], that in the dilute instanton gas appoximation
[27] instantons do not practically affect determination of
$\alpha_s(m^2_{\tau})$ and the Borel sum rules. Their effect,
however, appears to be considerable and strongly dependent on the
value of the instanton radius $\rho_c$ in the sum rules obtained
by integration over closed contours in the complex plane $s$ at
the radii of the contours $s < 2 GeV^2$.

\section{Sum rules for charmonium and gluonic
condensate.}

In this Section the charmonium sum rules are revisited.
 (In what follows I formulate the main results of
[28]).

Consider the polarization operator of charmed vector currents
\be
i\int dx \, e^{iqx} \left< \,TJ_\mu (x) J_\nu (0) \, \right>\,=\,
(\,q_\mu q_\nu - g_{\mu\nu}q^2 \,)\, \Pi(q^2) \; , \qquad J_\mu =
\bar{c} \gamma_\mu c \label{59} \ee The dispersion representation
for $\Pi(q^2)$ has the form
\be
R(s)\,=\,4\pi \, {\rm Im} \, \Pi(s+i0) \; , \qquad
\Pi(q^2)\,=\,{q^2\over 4\pi^2}\int_{4m^2}^\infty \,{R(s)\,ds\over
s(s-q^2)} \; , \label{60}\ee where $R(\infty) = 1$ in partonic
model. In approximation of infinitely narrow widths of resonances
$R(s)$ can be written as sums of contributions from resonances and
continuum
\be
\label{rexp} R(s)\,=\,{3 \, \pi \over Q_c^2 \, \alpha_{\rm
em}^2\!(s)}\, \sum_\psi m_\psi \Gamma_{\psi \to
ee}\,\delta(s-m_\psi^2) \,+\,\theta(s-s_0) \label{61}\ee where
$Q_c = 2/3$ is the charge of charmed quarks, $s_0$ - is the
continuum threshold (in what follows $\sqrt{s_0} = 4.6 GeV$), ~~
$\alpha(s)$ - is the running electromagnetic constant,~
$\alpha(m^2_{J/\psi}) = 1/133.6$ Following [1], to suppress the
contribution of higher states and continuum we will study the
polarization operator moments
\be
\label{momdef} M_n(Q^2)  \equiv {4\pi^2\over n!} \left( - {d\over
dQ^2}\right)^n\Pi(-Q^2)= \int_{4m^2}^\infty {R(s)\, ds\over
(s+Q^2)^{n+1}} \label{62}\ee According to (31) the experimental
values of moments are determind by the equality
\be
\label{momexp} M_n(Q^2)\,=\,{27\,\pi\over 4\, \alpha_{\rm
em}^2}\sum_{\psi=1}^6 {m_\psi\Gamma_{\psi\to ee}\over
(m_\psi^2+Q^2)^{n+1}} \,+\,{1\over n (s_0+Q^2)^n} \label{63}\ee It
is reasonable to consider the ratios of moments
$M_{n1}(Q^2)/M_{n2}(Q^2)$ from which the uncertainty due to error
in $\Gamma_{J/\psi \to ee}$ markedly falls out. Theoretical value
for $\Pi(q^2)$ is represented as a sum of perturbative and
nonperturbative contributions. It is convenient to express the
perturbative contribution through $R(s)$, making use of (30),
(32):
\be
R(s)\,=\,\sum_{n\ge 0} R^{(n)}(s,\mu^2)\, a^n(\mu^2) \label{64}\ee
where $a(\mu^2) = \alpha_s(\mu^2)/\pi$. Nowadays, three terms of
expansion in (34) are known: $R^{(0)}$ [29]~ $R^{(1)}$ [30], ~
$R^{(2)}$ [31].  They are represented as functions of quark
velocity $v = \sqrt{1 - 4m^2/s}$, ~ where $m$ - is the pole mass
of quark. Since they are cumbersome, I will not present them here.

Nonperturbative contributions into polarization operator have the form
(restricted by d=6 operators):
$$\Pi_{nonpert}(Q^2) = \frac{1}{(4m^2)^2} \langle 0\mid
\frac{\alpha_s}{\pi} G^2 \mid 0 \rangle [~f^{(0)}(z) +af^{(1)}
(z)~] + $$
\be
+\frac{1}{(4m^2)^3} g^3 f^{abc} \langle 0 \mid G^a_{\mu\nu}
G^b_{\nu\lambda} G^c_{\lambda \mu} \mid 0 \rangle F(z),~~
z=-\frac{Q^2}{4m^2}\label{65}\ee


\begin{figure}[tb]
\hspace{30mm} \epsfig{file=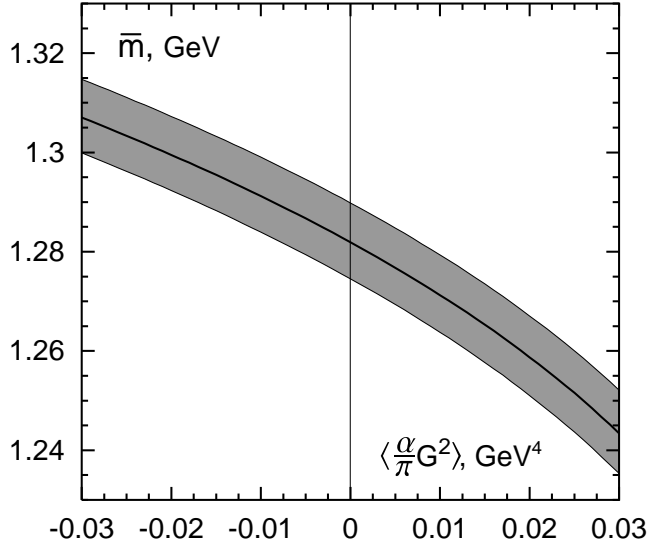, width=85mm} \caption{The
dependence of $\overline{m}(\overline{m})$ on $\langle 0 \vert
\alpha s/\pi)G^2 \vert 0 \rangle$ obtained at $n = 10$,~ $Q^2 =
0.98 \cdot 4m^2$ and $\alpha_s(Q^2 + \overline{m}^2)$. }
\end{figure}


\begin{figure}[tb]
\hspace{0mm} \epsfig{file=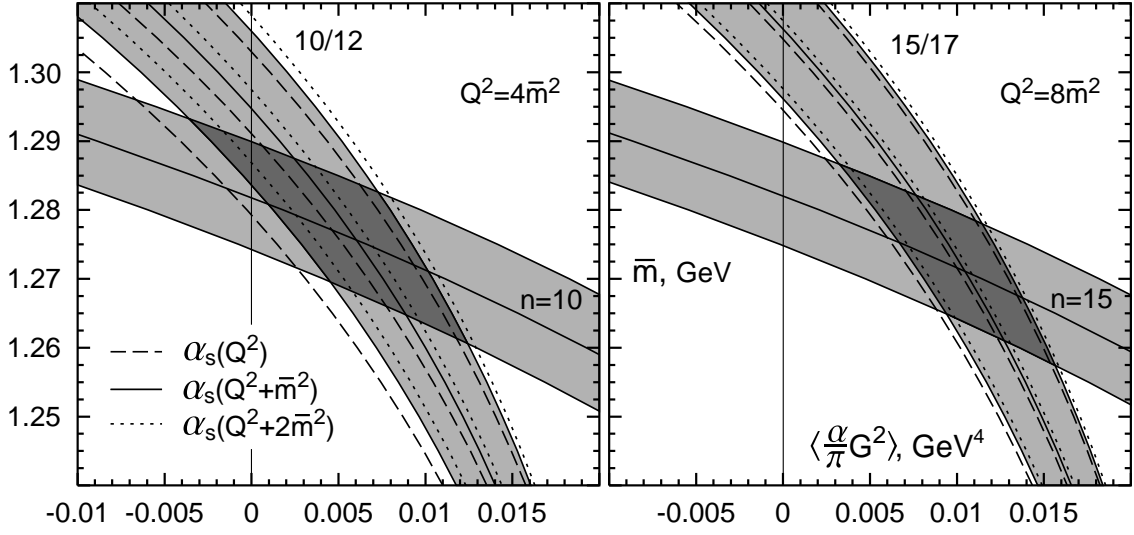, width=150mm} \caption{The
dependence of $\overline{m}(\overline{m})$ on $\langle 0 \vert
(\alpha s/\pi) G^2 \vert 0 \rangle$ obtained from the moments
(horizontal bands) and their ratios (vertical bands) at different
$\alpha_s$. The left-hand figure: $Q^2 = 4 \overline{m}^2$, ~ $n =
10$, ~ $M_{10}/M_{12}$; the right-hand figure -- $Q^2 = 8
\overline{m}^2$, ~ $n = 15$, ~ $M_{15}/M_{17}$. }
\end{figure}
\noindent
 Functions $f^{(0)}(z)$, ~ $f^{(1)}(z)$ and $F(z)$ were
calculated in [1], [32], [33], respectively. The use of the quark
pole mass is, however, inacceptable. So, it is reasonable to turn
 to $\overline{MS}$ mass $\overline{m}(\mu^2)$, taken at the
point $\mu^2 = \overline{m}^2$.  After turning to the
$\overline{MS}$ mass $\overline{m}(\overline{m}^2)$ we
\be
n=10 \; , \; Q^2=4{\bar m}^2:  \qquad {{\bar
M}^{(1)}\over {\bar M}^{(0)}}=0.045 \; , \qquad {{\bar M}^{(2)}\over {\bar
M}^{(0)}}=1.136 \; , \qquad {{\bar M}^{(G,1)}\over {\bar M}^{(G,0)}}=-1.673
\label{67} \ee

At $a \sim 0.1$ and at the ratios of moments given by (36) there
is a good reason to believe that the PT series well converges.
Such a good convergence holds (at $n > 5$) only in the case of
large enough $Q^2$, at $Q^2 = 0$  one does not succeed in finding
such $n$, that perturbative corrections, $\alpha_s$ corrections to
gluonic condensates and the term $\sim \langle G^3 \rangle$
contribution would be simultaneously small.

It is also necessary to choose the scale - normalization point
$\mu^2$ where $\alpha_s(\mu^2)$ is taken. In (34)~ $R(s)$ is a
physical value and cannot depend on $\mu^2$. Since, however, we
take into account in (34) only three terms, at unsuitable choice
of $\mu^2$ such $\mu^2$ dependence may arise due to neglected
terms. At large $Q^2$  the natural choice is $\mu^2 = Q^2$. It can
be thought that at $Q^2 = 0$ the reasonable scale is $\mu^2 =
\overline{m}^2$, though some numerical factor is not excluded in
this equality. That is why it is reasonable to take interpolation
form
\be
\mu^2 = Q^2+\overline{m}^2,\label{68}\ee but to check the
dependence of final results on a possible factor at
$\overline{m}^2$. Equalling theoretical value of some moment at
fixed  $Q^2$ (in the region where $M^{(1)}_n$ and $M^{(2)}_n$ are
small) to its experimental value one can find the dependence of
$\overline{m}$ on $\langle(\alpha_s/\pi)G^2 \rangle$ (neglecting
the terms $\sim \langle G^3 \rangle$). Such a dependence for $n =
10$ and $Q^2/4 m^2 = 0.98$ is presented in Fig.6.

To fix both $\overline{m}$ and $\langle(\alpha_s/\pi) G^2 \rangle$
one should, except for moments, take their ratios. Fig.7 shows the
value of $\overline{m}$ obtained from the moment $M_{10}$ and the
ratio $M_{10}/M_{12}$ at $Q^2 = 4 m^2$ and from the moment
$M_{15}$ and the ratio $M_{15}/M_{17}$ at $Q^2 = 8 m^2$. The best
values of masses of charmed quark and gluonic condensate are
obtained from Fig.7:
\be
{\bar m}({\bar m}^2)\,=\,1.275\pm 0.015 \, {\rm GeV} \; , \qquad
\left< {\alpha_s\over \pi} G^2\right>\,=\,0.009\pm 0.007 \, {\rm
GeV}^4 \label{69}\ee Up to now the corrections $\sim \langle G^3
\rangle$ were not taken into account. It appears that in the
region of $n$ and $Q^2$ used to find $\overline{m}$ and gluonic
condensate they are comparatively small and, practically, not
changing $\overline{m}$, increase $\langle
(\alpha_s/\pi)G^2\rangle$ by $10-20\%$ if the term $\sim \langle
G^3 \rangle$ is estimated according to instanton  gas model [34]
at $\rho_c = 0.5 fm$.

It should be noted that improvement of the accuracy of
$\Gamma_{J/\psi \to ee}$ would make it possible to precise the
value of gluonic condensate: the widths of horizontal bands in
fig.7 are determined mainly just by this error. In particular,
this, perhaps, would allow one to exclude the zero value of
gluonic condensate, that would be extremely important.
Unfortunately, eq.(38) does not allow one to do it for sure.
Diminution of theoretical error which determine the width  of
vertical bands seems to be less real.

\section{Conclusion}

In this report I compare the results of the recent precise
measurements of $\tau$-lepton hadronic decays [4]-[6] with QCD
predictions in the low energy region. The perturbative terms up to
$\alpha^3_s$ and the terms of the operator product expansion (OPE)
up to d=8 were taken into account. It is shown that QCD with the
account of OPE terms agrees with experiment up to $\sim 2\%$ at
the values of the complex Borel parameter $\vert M^2 \vert >
0.8-1.0 GeV^2$ in the left-hand semiplane of the complex plane. It
was found:\\ 1. The values of the QCD coupling constant
$\alpha_s(m^2_{\tau}) = 0.355 \pm 0.025$ from the total
probability of $\tau$-decays and $\alpha_s(m^2_{\tau}) = 0.330$
from the sum rules at low energies. (The latter value corresponds
to $\alpha_s(m^2_z) = 0.118$).\\ 2. The value of the quark
condensate square (assuming factorization) $$ \alpha_s \langle
\mid \overline{\psi}\psi \mid 0\rangle^2 = (2.25 \pm 0.70)\cdot
10^{-4}~\mbox{GeV}^6$$  and of quark-gluon condensate of d=8.\\ 3.
The value of gluonic condensate:

a) from the $\tau$-decay data: $$\langle 0\mid
\frac{\alpha_s}{\pi} G^2 \mid 0 \rangle = (0.006\pm
0.012)~\mbox{GeV}^4$$

b) from the sum rules for charmonium
 $$\langle 0\mid
\frac{\alpha_s}{\pi} G^2 \mid 0 \rangle = (0.009\pm 0.007)~
GeV^4$$ It is shown that the sum rules for charmonium are in
agreement with experiment when accounting for perturbative
corrections $\sim \alpha^2_s$ and for OPE terms proportional to
$\langle (\alpha_s/\pi) G^2 \rangle$ and to $\langle G^3 \rangle$.

The main conclusion is that in the range of low-energy phenomena
under consideration, perturbation theory and operator product
expansion are  in an excellent agreement with experiment starting
from $Q^2 \sim 1 GeV^2$.

I am deeply indebted to K.N.Zyablyuk who had made the main
calculations in papers [7,8,28], the results of which I used here.

This work was  supported by the grants CRDF RP2-2247,
INTAS-2000-587 and RFFI 00-02-17808.

\newpage


\begin{thebibliography}{99}
\bibitem{ab} M.A.Shifman, A.I.Vainshtein and V.I.Zakharov,
Nucl. Phys. {\bf 147}, 385, 448 (1979).
\bibitem{ac} B.L.Ioffe in: Proc. of the XXII International
Confertence  on High Energy Physics, Leipzig, 1984, v.2, p.176\\
B.L.Ioffe, Lectures in XXIII Cracow School of Physics, Acta
Physica Polonica {\bf B16}, 543 (1984).
\bibitem{ad} P.Colangelo and A.Khodjamirian  in: Handbook of QCD,
Boris Ioffe Festschrift, ed. by M.Shifman, World Scientific 2001,
v.3, p.1495. (1998).\bibitem{24} ALEPH Collaboration, R.Barate et
al., Eur.Phys.J.C {\bf 4}, 409 (1998).
\bibitem{25} OPAL Collaboration, K.Ackerstaff et al., Eur.Phys.J.C
{\bf 7}, 571 (1999); G.Abbiendi et al., ibid, {\bf 13}, 197
(2002).
\bibitem{26} CLEO Collaboration, S.J.Richichi et al., Phys.Rev.D
{\bf 60}, 112002 (1999). \bibitem{27} B.L.Ioffe, K.N.Zyablyuk,
Nucl.Phys.A {\bf 687}, 437 (2001). \bibitem{28} B.V.Geshkenbein,
B.L.Ioffe, K.N.Zyablyuk, Phys.Rev.D {\bf 64}, 093009 (2001).
\bibitem{29} A.Pich, Proc. of QCD94 Workshop, Monpellier, 1944;
Nucl.Phys.B (proc.Suppl) {\bf 39}, 396 (1995). \bibitem{30}
W.J.Marciano, A.Sirlin, Phys.Rev.Lett. {\bf 61}, 1815 (1988).
\bibitem{31} E.Braaten, Phys.Rev.Lett. {\bf 60}, 1606 (1988);
Phys.Rev.D {\bf 39}, 1458 (1989). \bibitem{32} S.Narison, A.Pich,
Phys.Lett.B {\bf 211}, 183 (1988).
\bibitem{33} F.Le Diberder, A.Pich, Phys.Lett.B {\bf 286}, 147
(1992). \bibitem{34} K.G.Chetyrkin, A.L.Kataev, F.V.Tkachov,
Phys.Lett.B {\bf 85}, 277 (1979); M.Dine, J.Sapirshtein,
Phys.Rev.Lett. {\bf 43} 668 (1979); W.Celmaster, R.Gonsalves,
ibid, {\bf 44}, 560 (1980). \bibitem{35} L.R.Surgaladze,
M.A.Samuel, Phys.Rev.Lett. {\bf 66}, 560 (1991);\\ S.G.Goryshny,
A.L.Kataev, S.A.Larin, Phys.Lett.B {\bf 259}, 144 (1991).
\bibitem{36}
A.L.Kataev, V.V.Starshenko, Mod.Phys.Lett.A {\bf 10}, 235 (1995).
\bibitem{37} O.V.Tarasov, A.A.Vladimirov, A.Yu.Zharkov,
Phys.Lett.B  {\bf 93}, 429 (1980); S.A.Larin, J.A.M.Vermaseren,
ibid, {\bf 303}, 334 (1993). \bibitem{38} T.van Ritbergen,
J.A.M.Vermaseren, S.A.Larin, Phys.Lett.B {\bf 400}, 379 (1997).
\bibitem{39} K.G.Chetyrkin, S.G.Gorishny, V.P.Spiridonov,
Phys.Lett.B {\bf 160}, 149 (1985). \bibitem{40} L.-E.Adam,
K.G.Chetyrkin, Phys.Lett.B {\bf 329}, 129 (1994). \bibitem{41}
E.Braaten, C.S.Lee, Phys.Rev.D {\bf 42}, 3888 (1990).
\bibitem{42}
K.Hagiwara et al., Particle Data Groop, Phys.Rev.D {\bf 66},
010001 (2002).\bibitem{43} ALEPH Collaboration, R.Barate et al.,
Eur.Phys.J.C {\bf 11}, 599 (1999).
\bibitem{44} OPAL Collaboration, G.Abbiendi et al., Eur.Phys.J.C
{\bf 19}, 653 (2001).
\bibitem{cj} V.A.Novikov, M.A.Shifman, A.I.Vainstein and
V.I.Zakharov, Nucl.Phys. B {\bf 249}, 445 (1985).
\bibitem{ck} M.A.Shifman, {\it Lecture at 1997 Yukawa
International Seminar}, Kyoto, 1997, Suppl.Prog.Theor.Phys., 1998,
Vol.131, p.1.
\bibitem{47} T.Shafer, E.V.Shuryak, Rev.Mod.Phys. {\bf 70}, 323
(1998).
\bibitem{48} B.L.Ioffe, K.N.Zyablyuk, hep-ph/0207183.
\bibitem{49} V.B.Berestetsky, I.Ya.Pomeranchuk, JETP {\bf 29}, 864
(1955).
\bibitem{50} J.Schwinger, Particles, Sources, Fields, Addison-Wesley
Publ.,
1973, V.2. \bibitem{51} A.H.Hoang, J.H.Kuhn, T.Teubner,
Nucl.Phys.B {\bf 452}, 173 (1995);\\ K.G.Chetyrkin, J.H.Kuhn,
M.Steinhauser, Nucl.Phys.B {\bf 482}, 213 (1996);\\ K.G.Chetyrkin
et al., Nucl.Phys.B {\bf 503}, 339 (1997);\\ K.G.Chetyrkin et al.,
Eur.Phys.J.C {\bf 2}, 137 (1998).
\bibitem{52} D.J.Broadhurst et al., Phys.Lett.B {\bf 329}, 103
(1994). \bibitem{53} S.N.Nikolaev, A.V.Radyushkin, Yad.Fiz. {\bf
39}, 147 (1984).
\bibitem{fg} V.A.Novikov, M.A.Shifman, A.I.Vainshtein,
V.I.Zakharov, Phys.Let. {\bf B86}, 347 (1979).

\end{thebibliography}
\end{document}